\documentclass[12pt,preprint]{aastex}
\usepackage{psfig}
\begin{document}
\title{A Search for Stars of Very Low Metal Abundance. V.
Photoelectric $UBV$ Photometry of Metal-Weak Candidates from the Northern HK
Survey}
\author{P.Bonifacio\altaffilmark{1}, S. Monai\altaffilmark{1}, \& T.C.
Beers\altaffilmark{2}}

\altaffiltext{1}{Osservatorio Astronomico di Trieste, Via G.B. Tiepolo 11,
34131--Trieste, Italy;\\ email: bonifaci@ts.astro.it, monai@ts.astro.it} 
\altaffiltext{2}{Michigan State University, Department of Physics \& Astronomy, E.
Lansing, MI 48824;\\ email: beers@pa.msu.edu}

\slugcomment{To Appear in the Astronomical Journal, October 2000}
\begin{abstract}

We report photoelectric $UBV$ data for 268 metal-poor candidates chosen from
the northern HK objective-prism/interference-filter survey of Beers and
colleagues.  Over 40 \% of the stars have been observed on more than one
night, and most have at least several sets of photometric measurements.
Reddening estimates, preliminary spectroscopic measurements of abundance, and
type classifications are reported.

\end{abstract}

\section{Introduction}

The HK objective-prism/interference-filter survey, started in 1978 by Preston
\& Shectman (see Beers, Preston, \& Schectman 1985; Beers, Preston, \& Shectman
1992), has proven to be a most effective method for the discovery of relatively
bright metal-deficient stars in the thick disk and halo of the Galaxy, and is
the primary modern source of stars known with [Fe/H] $ < -3.0$.  Such stars are
of crucial importance, as they provide fundamental insight on the
nucleosynthetic history of the early Galaxy.

The spectroscopic and photometric follow-up of candidate metal-poor stars from
the HK survey has become a large international effort, with observational
programs being conducted by colleagues with access to both northern and
southern hemisphere telescopes.  Besides the identification of extremely
metal--poor stars, the HK survey is providing a large number of stars of
intermediate metallicity ($-0.5 \le {\rm [Fe/H]} \le -2.0$) -- stars that are 
useful for the study of the thick disk, the metal-weak thick disk (MWTD), 
and of the thick disk to halo transition.

Since the HK survey is kinematically unbiased in its selection criteria, it is
also an ideal resource for studies of the kinematics and dynamics of the halo
and thick-disk populations.  Many of the brighter HK survey candidates have
available proper motion measurements, and, when combined with accurate
(photometric, at present) distance estimates and radial velocities, yield a
large sample of stars with full space motions (e.g., Beers et al. 2000).  Based
on such detailed information, Chiba \& Beers (2000) explore a number of
important issues related to the formation and evolution of the Galaxy, and show
how the kinematics of local samples can be used to provide constraints on the
low-metallicity cutoff of the MWTD population ([Fe/H] $\sim -2.0$), the degree
of flattening in the thick-disk and halo populations, and the variation of
flattening with Galactocentric distance, the existence (or not) of the supposed
``counter-rotating high-halo population,'' measurements of velocity gradients
in the thick disk and halo, and searches for kinematic substructure such as
reported by Helmi et al. (1999).

A large campaign to obtain medium-resolution spectroscopic observations of
additional HK survey metal-poor candidates has been underway for roughly the
past seven years, and the total number of stars with suitable spectra for
abundance and radial velocity measurement is now on the order of $N \sim 5000$
(see Beers 1999 for a summary).  Unfortunately, the photometric follow-up
campaign has been lagging behind somewhat, especially in the northern
hemisphere, due to the lack of suitable telescopes (and sites) for carrying out
the observations (a problem that is getting worse, not better, as time
passes).   The bulk of the extant northern $UBV$ data comes from a study of 300
HK candidates by Doinidis \& Beers (1990, 1991).  Str\"omgren photometry of a
sample of over 500 northern HK survey stars has been recently reported by
Anthony-Twarog et al.  (2000).  Additional Str\"omgren photometry of 140
northern HK metal-poor candidates has been obtained by Schuster et al. (2000).
In the southern hemisphere the situation is somewhat better, though additional
work is still required.  Preston, Shectman, \& Beers (1991) obtained $UBV$ for
a sample of $\sim 1800$ southern HK survey stars, though most were field
horizontal-branch (hereafter, FHB) stars rather than cooler metal-poor
candidates.  $UBV$ measurements for a large sample ($N \sim 2600$)
of southern HK survey candidate metal-poor stars have been recently reported by
Norris, Ryan, \& Beers (1999).  Schuster et al.  (1996) report Str\"omgren
photometry for a small ($N \sim 90$) sample of metal-poor HK survey stars,
including both northern and southern candidates.

The photometry is important mainly for two reasons -- (1) to improve the
metallicity estimates of the stars, making use of the de-reddened $(B-V)_0$
color (or its approximation from the Str\"omgren $(b-y)_0$ index) and the
calibration procedure described by Beers et al. (1999), and (2) for obtaining
photometric distance estimates.  For the latter purpose it is important to also
have the $(U-B)_0$ color, since it enables one to discriminate (in most, but
not all cases) between FHB stars and main-sequence gravity stars near the
turnoff of an old stellar population (hereafter, TO) in the color range $0.3
\le (B-V)_0 \le 0.5$.

We thus decided to initiate a northern photometric campaign with a small (91
cm) telescope (somewhat undersized for photoelectric follow-up of
our targets, which have apparent magnitudes mainly in the range $12.5 < V <
14.5$) for which we were granted substantial amounts of observing time over an
extended period.  The results of the first year of this campaign (from August
1998 to August 1999) are quite satisfactory.  We obtained data on 22 nights
out of 30 allocated, and here we report $UBV$ for a total of 268 
stars.  Roughly 40 \% of these were observed on more than one night.
For most stars observations on different nights were consistent and their
average provides accurate magnitudes and colors, but for a small number of
stars they were not.  We also present data for these stars, flagging them as
``suspected variables.''  All but a handful of our stars already have
medium-resolution spectroscopy available, so the reported photometry can be
put to immediate use.

\section{Observations and Data Reduction}

The observations were carried out with the 91 cm telescope of the Osservatorio
Astrofisico di Catania at the M.G. Fracastoro mountain station on Mt. Etna
during 4 runs in August 1998, October 1998, May 1999, and August 1999.  A log
of the observations is given in Table 1, along with estimates of the typical
residuals (relative to the photometric standard stars) that were achieved over
the course of the run.  All observations were performed with a single-channel
photoelectric photometer mounted at the Cassegrain focus.  The photomultiplier
tube was a EMI 9893QA/350, which mimics a S20 spectral response, used along
with Johnson $UBV$ filters.  For all the observations we employed diaphragm
\#2, corresponding to a $21\farcs 7$ diameter on the sky.  Calibrations
were obtained by observations of typically 10--15 standard stars each night,
taken from the lists of Landolt (1983, 1992).  The calibration scheme followed
the method outlined by Harris, Fitzgerald, \& Reed (1981), retaining only
linear terms in the colors.  This method allows one to tie several different
nights into a single calibration, and hence recover usable data for ``partial''
nights during which few standard stars were observed due to changes in the
weather conditions.

Given its location at the foot of an active volcano, the Serra La Nave
Observatory is in a somewhat particular situation.  Typically the wind blows
from west to east, thus keeping smoke from the volcano away from the
Observatory.  When the wind blows from east to west the smoke moves in the
direction of the Observatory, and it is necessary to close the dome to avoid
the dust damaging the mirror.  In addition, there are many intermediate
situations in which, either due to a slightly different direction of the wind
or to extra activity of the volcano, the measured extinction is markedly
different from the ``typical'' values -- these nights may well be photometric,
however they must be calibrated on their own and may not be tied to other
nights. 

Our reduction scheme began by performing a calibration of each night
individually, enabling rejection of ``bad'' standards (usually easily
identified as outliers).  Next we compared the derived extinctions of the
different nights.  Nights with highly deviant extinctions were left with their
individual calibration, while the others were tied together with a single
``multi-night'' calibration.  On a few ``odd'' nights it was necessary to
introduce a linear term in UT into the extinction equations.  In Table 1 we
indicate the nights that have been treated with the ``multi-night'' reduction
scheme (indicated by an ``m''in the fifth column), and the ones that were
calibrated individually (indicated by an``s'' in the fifth column). A ``t'' in
the last column denotes nights for which a time-dependent extinction has been
assumed.  No attempt was made to tie observations taken on different runs into
a single calibration, because the seasonal change in the extinction is likely
to add uncertainty to the resulting calibration.

Once all the data had been calibrated we averaged together the available
observations of each star.  Some stars were observed on only one night, even
though several times, while others were observed  on several different nights.
These stars broadly divided into two groups: stars whose standard deviations in
the derived magnitudes were less than or equal to the  expected
error and others for which they were larger.  We define the first group as
``stars that display no variability,'' and list their photometric information
in Table 2.  Column (1) of this table lists the star name. Columns (2) and (3)
are the (1950) epoch coordinates (accurate to several arc-seconds), and columns
(4) and (5) are the 
(2000) epoch coordinates.
Columns (6) and (7) are
the corresponding Galactic coordinates of each star.  Columns
(8)-(13) list the measured $V$, $B-V$, and $U-B$, and their associated
one-sigma errors.  Columns (14) and (15) list the numbers of independent nights
on which the star was observed, and the total number of (useful) observations
per star that were employed, respectively.  The final two columns of this
table list two alternative estimates of reddening -- column (16) is that
derived from the new reddening maps of Schlegel, Finkbeiner, \& Davis (1998),
while column (17) is that obtained from the maps of Burstein \& Heiles (1982).
We describe our choice of reddening below.

The second set of stars are reported  as ``suspected variables.''  As a
dividing line we chose the standard deviation in the $V$ magnitude: if
$\sigma_V > 0.08$ mag we consider the star to be a ``suspected variable.'' This
division is to some extent arbitrary and the status of ``suspected variables''
should mainly be used as a warning flag.  For all the ``suspected variables''
that were observed on at least three different nights we computed a Lomb
normalized periodogram, using the routine {\tt period} of Press et al (1992).
Only the star BS 16542--0002 displayed peaks with a significance above 95\% --
we therefore consider this as the only confirmed variable of our present
sample. This star exhibits colors ($(B-V)_0 = 0.70$) that exclude it from being
an RR Lyrae variable, so it merits additional attention to establish the source
of its apparent variation.  The periodogram analysis is optimized for periodic
variables, so that any irregular or semi-regular variables are likely to go
undetected by this criterion.  In Table 3, we list the mean magnitudes and
colors, and their standard deviations, for ``suspected'' variables, following
the same column assignments as in Table 2.

Note that a number of objective-prism plates in the HK survey overlap with one
another, hence there are several instances of candidates with multiple names.
Table 4 summarizes the multiple identifications for the HK survey stars in our
present sample.

\section{External Comparisons}

There are 87 stars in common between the present paper and the paper of
Anthony-Twarog et al. (2000), which provides $uvbyCa$ photometry.  Our $V$
magnitudes may be directly compared to theirs, which were obtained by
calibrating the $y$ filter to the Johnson $V$ system. In two cases the
difference is large enough that we may be almost certain the we did not observe
the same star.  In the case of BS~17570--0057 our measurements indicate
$V=13.40, B-V=0.43$, while Anthony-Twarog et al. give $V=9.972, B-V\approx 1.35
\times b-y = 0.87$.  The Guide Star Catalog lists an estimated $V$ magnitude
for this star of $V=13.4$, and our medium-resolution spectra provide an
estimate of $B-V \sim 0.47$, based on the calibration of the HP2 Balmer-line 
index described in Beers et al. (1999).  We observed the star on two different
nights for a total of four measurements, and the measurements were all
consistent with one another.  In the case of BS~17583--0012 the situation is
less clear.  We observed the star on one night only, and obtain $V=12.45,
B-V=0.54$, while Anthony-Twarog et al.  give $V=11.497$, and from their
reported $b-y$ color we estimate $B-V \approx 0.83$.  For this star the Guide
Star Catalog lists $V=12.1$, and from the HP2 index we estimate $B-V \sim
0.39$.  Hence it is possible that we observed different stars, possibly neither
of which was the correct one, however it is also possible that the star is
variable.  We exclude these two stars from the comparison.  Future measurements
will permit us to establish the correct photometry for both stars and to check
for variability.

The top panel of Figure 1 shows a comparison of the $V$ magnitudes reported by
Anthony-Twarog et al., $V\;(AT)$, with our measurements for the remaining 85
stars--the agreement is quite satisfactory.  We estimate the difference in
central location (``mean'') between these two samples, and the scale
(``standard deviation'') of their differences, using the robust and resistant
biweight estimators described by Beers, Flynn, \& Gebhardt (1990).  We obtain
$C_{BI} (\Delta V) = -0.003$ magnitudes, and $S_{BI} (\Delta V) = 0.053$
magnitudes.  The biweight scale may be considered as a conservative estimate of
the external accuracy of our photometry, if we assume that the $V$ magnitudes
of Anthony-Twarog et al. are error-free, and the scatter in $\Delta V$ is due
solely to the errors in our photometry.  A more realistic error estimate is
obtained by assuming that the external errors of the photometry of both groups
is the same and dividing the above value by $\sqrt 2$, obtaining 0.04
magnitudes, which is on the order of the reliability of the reddening
corrections discussed below.

The lower panel of Figure 1 shows the comparison of
the estimated $B-V$ color of
Anthony-Twarog et al., derived from the standard relation $B-V_{est} = 1.35
\times b-y$, with our measured $B-V$ colors.  The agreement is generally good,
with the possible exception of the bluest stars, which could of course 
include bona-fide variables such as RR Lyraes.  The biweight estimate of
central location of the difference in color $\Delta (B-V) = (B-V)\; (our) -
(B-V)_{est}\; (AT)$) is  $C_{BI} = -0.02$ magnitudes.  The
biweight scale estimate of the difference in colors is $S_{BI} = 0.06$
magnitudes.  Apportioning the errors equally between the two studies suggests
an external accuracy of our $B-V$ values of $\sim 0.04$ magnitudes.  This is
perhaps a bit too pessimistic, since we are comparing two different photometric
systems (Str\" omgren versus Johnson), and have assumed that the transformation
equations are both error free and apply equally well over the range of colors
and abundances of our program stars.

\section{Adopted Reddenings, Preliminary Abundances, and Stellar Classifications}

We now consider the nature of our target stars, following procedures described
by Beers et al. (1999), slightly modified as described below.

\subsection {Reddening Estimates}

We initially adopted the Schlegel et al. (1998) estimates of reddening from
Tables 2 and 3, which have superior spatial resolution and are thought to have
a better-determined zero point than the Burstein \& Heiles (1982) maps.
However, Arce \& Goodman (1999) caution that the Schlegel et al. map
may overestimate the reddening values when the color excess $E\;(B-V)_S$
exceeds about 0.15 magnitudes.  Our own independent tests suggest that
this problem may extend to even lower color excesses, on the order of
$E\;(B-V)_S = 0.10$ magnitudes.  Hence, we have adopted a slight revision of
the Schlegel et al. reddening estimates, according to the following:

\begin{equation}
\begin{array}{lclcl}
E\;(B-V)_A & = & E\;(B-V)_S &\;\;\;\;\; & E\;(B-V)_S \le 0.10\\ [.25in]
E\;(B-V)_A & = & 0.10 + 0.65 \times [E\;(B-V)_S - 0.10] &\;\;\;\;\;&  E\;(B-V)_S > 0.10
\end{array}
\end{equation}

\noindent where $E\;(B-V)_A$ indicates the adopted reddening estimate.
We note that for $E\;(B-V)_S \ge 0.15$ this approximately reproduces the
30\%--50\% reddening reduction recommended by Arce \& Goodman.  To account for
stars that are located within the reddening layer, assumed to have a scale
height $h = 125$ pc,  the reddening to a given star at distance $D$ is reduced
compared to the total reddening by a factor $[1-\exp(-|D\; \sin\; b|/h)]$,
where $b$ 
is the Galactic latitude.  The distance is estimated from $M_V ~~vs. ~~
(B-V)_0$ relations, as described in Beers et al. (2000).  The procedure must be
iterated, because both $V_0$ (and therefore $D$) and  $(B-V)_0$ depend on the
$E\;(B-V)_A$.  Since the $M_V ~~vs. ~~ (B-V)_0$ relation depends on
metallicity, as well as on the classification of the star, at each step of the
iteration the metallicity is recomputed and the classifications redetermined
with the current estimates of $(B-V)_0$ and $(U-B)_0$, so that at the end we
obtain consistent estimates of $E\;(B-V)_A$, [Fe/H], and $D$.  The procedure is
not without shortcomings, as in some regions of the color and abundance space
there is little or no power in discrimination of type classifications.  This
classification degeneracy adds some uncertainty, as discussed below.
 
The final adopted estimate of reddening is listed in column (2) of Tables 5 and
6.  Column (3) lists the resulting extinction-corrected $V_0$ magnitudes, using
$A_V = 3.1\;E\;(B-V)_A$.  Reddening-corrected colors, obtained from $(B-V)_0 =
B-V - E\;(B-V)_A$, and $(U-B)_0 = U-B - 0.72\;E\;(B-V)_A$ are listed in columns
(4) and (5), respectively.

\subsection{Abundance Estimates}

Medium-resolution spectroscopy was obtained for the majority of our program
stars with either the 2.1m telescope at Kitt Peak National
Observatory, using the Goldcam spectrograph, or the 2.5m Isaac Newton Telescope
on La Palma, using the Intermediate Dispersion Spectrograph.  For a few stars
we have combined estimates obtained from other spectroscopic follow-up
campaigns.  Details of the acquisition of this data are discussed  in
Beers et al. (1999) and Allende Prieto et al. (2000), and will not be repeated
here.  The full set of medium-resolution spectroscopic results for HK survey
stars from the recent campaigns will be presented in subsequent papers.

Measurements of the KP index (a pseudo-equivalent width of the CaII K line)
obtained from the spectra were used to obtain the metallicity estimate
[Fe/H]$_{\rm K3}$, following the procedures outlined in Beers et al., employing
the de-reddened $(B-V)_0$ colors described above.  The resulting abundance
estimates, and their associated one-sigma errors, are listed in column (6) and
(7) of Tables 5 and 6, respectively. For a few of the redder stars, inspection
of the spectra makes it clear that the CaII K line exhibits the presence of
emission in its core, likely compromising the KP index, and hence, the
estimated metallicity.  We have refrained from quoting abundances, or making
type classifications, for these stars, and have indicated them with the
designation ``CORE EM'' in column (9) of the tables.

For each star with an available spectrum we have also made independent
estimates of the expected $(B-V)_0$ color, derived from the calibration of this
color with the HP2 Balmer-line index described by Beers et al (1999).  In a
number of cases, this calibration yielded predicted de-reddened colors that
differed from our measured colors by larger than 0.15 magnitudes (2--3 times
the expected error in the HP2 calibration) -- the abundances of these stars are
indicated as somewhat uncertain by an appended ``:'' in column (6) of
Tables 5 and 6.  Stars classified as TO/FHB or FHB/A in column (8) of Tables 5
and 6 (as discussed below) also have some uncertainty in their adopted
metallicities, which we indicate by an appended ":" to their abundance
estimates.  A few stars in our present sample have colors that extend beyond
the nominal region of the Beers et al. metallicity calibration ($0.30 \le
(B-V)_0 \le 1.2$).  These stars are indicated in Column (9) of the tables with
the designation ``BLUE'' or ``RED'', and their abundance estimates have an
appended ``::'' to indicate the additional uncertainty.   For the two stars
with colors blueward of $(B-V)_0 = 0.0$ no abundance estimates is given.

\subsection{Stellar Classifications}

Classification of the program stars follows the procedures described in Beers
et al. (1999), with the following caveat.  The original classification
scheme (dating back to Beers et al. 1985) was designed for application to stars
of very low metallicity ([Fe/H] $< -2.0$).  However, since the HK follow-up 
now includes large numbers of stars with higher abundances, the procedure by
which the ``split'' between FHB and TO stars is made has to be modified.  In
the original scheme, stars in the color range $0.3 \le (B-V)_0 \le 0.5$, which
includes both FHB and TO stars, were assigned classifications based on the
simple criteria:

\begin{eqnarray*}
 TO:\;\; & \; & (U-B)_0 \le -0.1 \\ [0.25in]
FHB:\;\; & \; & (U-B)_0 > -0.1 
\end{eqnarray*}

\noindent This simple criteria is not adequate for stars with abundances [Fe/H]
$> -2.0$.

Figure 2 shows the ``limiting'' $(U-B)_0$ colors as a function of
metallicity, [Fe/H], as predicted from the Revised Yale Isochrones (Green 1988;
King, Demarque, \& Green 1988) for (a) stars with surface gravities $\log\; g
\le 3.0$ (appropriate for FHB stars in this color range--see
Wilhelm, Beers, \& Gray 1999) and (b) stars with $\log\; g \ge 4.0$
(appropriate for TO stars in this color range).  For the present
exercise, we make the classification assignments by comparison with these
limiting lines (broadened slightly to include possible reddening errors on the
order of 0.03 magnitudes).  Fits to these lines are well produced by the
following quadratic relations:

\begin{eqnarray*}
 TO:\;\; & (U-B)_0 (lim) =  & 0.080 + 0.207\; {\rm [Fe/H]} + 0.038\;{\rm [Fe/H]}^2 + 0.02 \\ [0.25in]
FHB:\;\; & (U-B)_0 (lim) =  & 0.105 + 0.130\; {\rm [Fe/H]} + 0.020\;{\rm [Fe/H]}^2 - 0.02
\end{eqnarray*}

The 0.02 constant that is added or subtracted in the above relations is the
correction term for the reddening error in $(U-B)_0$.  Stars with $(U-B)_0$
colors {\it above} the TO line (i.e., at lower values of $(U-B)_0$) are
classified as TO, while those {\it below} the FHB line (i.e., at higher values
of $(U-B)_0$) are classified as FHB.  Stars of intermediate surface gravities
fall between the lines (as do some stars as a result of errors in their
metallicities or colors), and their classifications are less certain.  Since we
expect MOST of the stars in this color regime to be metal-rich TO stars
rather than metal-poor FHB stars, for stars between the two limiting lines we
assign the classification TO/FHB (indicating that TO is our preference in the
classification, and is the one used for distance determination in the
assignment of reddening).  In most cases accurate photometry should produce a
reliable assignment, which is listed (along with classifications for other
types carried out following the Beers et al. 1999 procedures) in column (8) of
Tables 5 and 6.  Note that for stars with colors blueward of $(B-V)_0 = 0.30$
we have adopted a classification of FHB/A, and do not attempt to refine this
assignment further at present (however, see Wilhelm et al. 1999 for a
description of how this might be done).  For the two stars with colors blueward
of $(B-V)_0 = 0.0$ no classification is given.

\section {Discussion}

Figure 3 is a two-color diagram, $(U-B)_0 \; vs. \; (B-V)_0$, for the stars in
our present sample not listed as suspected  variables, over the
color range for which reliable abundance estimates could be obtained.  A
solar-abundance main-sequence line, estimated from the Revised Yale Isochrones,
is shown for comparison.  

All of our program stars have available radial velocities (accurate to $\sim
7-10$ km~s$^{-1}$), and many of the brighter ones have proper motions available
from a number of sources.  We intend to expand our photometric program 
of the HK survey stars (especially those with available proper motions) and
thereby supplement the basic data with which we can carry out a detailed
investigation of the kinematics of thick disk and halo stars.

\acknowledgments

We are grateful to Prof. S. Catalano for generously allocating time to such a
long-term project.  Special thanks are due to G. Carbonaro, A. Di Stefano, G.
Occhipinti, M. Puleo, and S. Sciuto for their assistance during the
observations.  We wish to thank E. Marilli and C. Lo Presti for providing the
source code of the PHOT reduction package with which we implemented the
multi-night reduction scheme.  Finally we thank T. Valente for carefully
preparing the many finding charts needed for the observations.  This work was
supported in part by NATO grant 950875.  T.C.B also acknowledges
partial support from grant AST 95-29454 from the National Science Foundation.

\clearpage

\begin{deluxetable}{rrrrcc}
\tabletypesize{\scriptsize}
\tablewidth{0pt}
\tablenum{1}
\tablecaption{Log of Observations}
\tablehead{
\colhead{Date} & \colhead{$\sigma_V$} & \colhead{$\sigma_{B-V}$} &
\colhead{$\sigma_{U-B}$}& \colhead{r\tablenotemark{a}} & \colhead{Comment}
}
\startdata
27/08/1998  &   0.012  &   0.015 & 0.013 & s &\nl
28/08/1998  &   0.013  &   0.007 & 0.012 & s & \nl
\nl
15/10/1998  &          &         &       & m & \nl
16/10/1998  &          &         &       & m & \nl
18/10/1998  &          &         &       & m & \nl
21/10/1998  &   0.008  &   0.008 & 0.010 & m & \nl
\nl
22/10/1998  &   0.024  &   0.012 & 0.010 & s &  \nl
\nl
13/05/1999  &           &        &       & m &  \nl
15/05/1999  &           &        &       & m &  \nl
16/05/1999  &           &        &       & m &  \nl
18/05/1999  &           &        &       & m &  \nl
20/05/1999   &   0.028  &   0.014 & 0.018 & m & \nl
\nl
5/08/1999   &           &        &       &   m & \nl
6/08/1999   &           &        &       &   m & \nl
8/08/1999   &           &        &       &   m & \nl
11/08/1999  &           &        &       &   m & \nl
13/08/1999  &           &        &       &   m & \nl
14/08/1999  &           &        &       &   m & \nl
15/08/1999 &    0.017  &    0.013 &  0.022&  m & \nl
\nl
7/08/1999  &    0.031  &    0.012 &  0.014 & s & \nl
\nl
9/08/1999  &    0.015  &    0.015 &  0.028 & s & t\nl
\nl
10/08/1999 &    0.010  &    0.015 &  0.021 & s & t\nl

\enddata
\tablenotetext{a}{reduction type: m=multi--night; s=single-night\hfill}
\end{deluxetable}

\clearpage

\begin{deluxetable}{lccccrcrccccccccc}
\rotate
\tabletypesize{\scriptsize}
\tablecolumns{17}
\renewcommand{\tabcolsep}{4pt}
\renewcommand{\arraystretch}{.6} 
\tablewidth{0pt}
\tablenum{2}
\tablecaption{Photometric Data For Program Stars \label{table2}}
\tablehead{
\colhead{Star}  & \multicolumn{2}{c}{~~RA~~~\phantom{1950}~~~DEC} & \multicolumn{2}{c}{~~RA~~~\phantom{2000}~~~DEC}
& \colhead{$l\;\;$} &
\colhead{$b$} &  \colhead{$V\;$} & \colhead{$\sigma_V$} & \colhead{$B-V$} &
\colhead{$\sigma_{B-V}$} & \colhead{$U-B$} & \colhead{$\sigma_{U-B}$} &\colhead{$N$} & \colhead{$N$} &
 \multicolumn{2}{c}{$E(B-V)$} \\
\colhead{} &
\multicolumn{2}{c}{1950} &
\multicolumn{2}{c}{2000} &
\colhead{} &
\colhead{} &
\colhead{} &
\colhead{} &
\colhead{} &
\colhead{} &
\colhead{} &
\colhead{} &
\colhead{$_{ni}$} &
\colhead{$_{ob}$} &
\colhead{S} &
\colhead{B} \\ 
\colhead{(1)} &
\colhead{(2)} &
\colhead{(3)} &
\colhead{(4)} &
\colhead{(5)} &
\colhead{(6)} &
\colhead{(7)} &
\colhead{(8)} &
\colhead{(9)} &
\colhead{(10)} &
\colhead{(11)} &
\colhead{(12)} &
%\colhead{(13)} &
\multicolumn{3}{l}{(13)~~(14)~(15)} &
%\colhead{(15)} & 
\colhead{(16)} & 
\colhead{(17)} 
}
\startdata
BS 15622$-$0053  & 12 51 07.7 & $+$25 21 58 & 12 53 34.3 & $+$25 05 42 & 316.3 &  \phs 87.9 &  13.59 &   0.01 &   0.74 &   0.03 &  \phs 0.38 &   0.06 &   1 &  2 &  0.01 & 0.02\\
BS 15622$-$0055  & 12 48 55.6 & $+$27 11 30 & 12 51 21.9 & $+$26 55 12 & 298.5 &  \phs 89.8 &  14.05 &   0.03 &   0.40 &   0.02 & $-$0.17    &   0.01 &   1 &  2 &  0.01 & 0.00\\
BS 15624$-$0040  & 16 15 27.3 & $+$46 11 08 & 16 16 58.5 & $+$46 03 49 &  72.2 &  \phs 45.5 &  13.57 &   0.03 &   0.41 &   0.02 & $-$0.22    &   0.02 &   3 &  6 &  0.01 & 0.00\\
BS 16034$-$0101  & 15 48 50.0 & $+$54 49 36 & 15 50 04.1 & $+$54 40 35 &  85.9 &  \phs 47.6 &  12.36 &   0.01 &   0.43 &   0.01 & $-$0.04    &   0.01 &   1 &  2 &  0.01 & 0.00\\
BS 16034$-$0115  & 15 59 04.4 & $+$57 26 38 & 16 00 07.9 & $+$57 18 15 &  88.7 &  \phs 45.3 &  11.98 &   0.02 &   0.37 &   0.01 & $-$0.26    &   0.01 &   2 &  4 &  0.02 & 0.00\\
BS 16076$-$0006  & 12 45 54.2 & $+$21 13 02 & 12 48 22.6 & $+$20 56 41 & 296.3 &  \phs 83.8 &  13.36 &   0.01 &   0.52 &   0.02 & $-$0.19    &   0.02 &   1 &  4 &  0.03 & 0.03\\
BS 16076$-$0012  & 12 49 25.2 & $+$18 02 29 & 12 51 54.2 & $+$17 46 11 & 303.6 &  \phs 80.6 &  13.76 &   0.05 &   0.47 &   0.02 & $-$0.26    &   0.03 &   1 &  2 &  0.02 & 0.01\\
BS 16076$-$0040  & 12 54 47.9 & $+$22 49 28 & 12 57 14.8 & $+$22 33 16 & 319.3 &  \phs 85.2 &  12.00 &   0.01 &   0.40 &   0.01 & $-$0.27    &   0.03 &   1 &  2 &  0.04 & 0.03\\
\enddata  
\end{deluxetable}

\clearpage

\begin{deluxetable}{lccccrcrccccccccc}
\rotate
\tabletypesize{\scriptsize}
\tablecolumns{17}
\renewcommand{\tabcolsep}{4pt}
\renewcommand{\arraystretch}{.6}
\tablewidth{0pt}
\tablenum{3}
\tablecaption{Photometric Data for Suspected Variable Stars \label{table3}}
\tablehead{
\colhead{Star}  & \multicolumn{2}{c}{~~RA~~~\phantom{1950}~~~DEC} & \multicolumn
{2}{c}{~~RA~~~\phantom{2000}~~~DEC}
& \colhead{$l\;\;$} &
\colhead{$b$} &  \colhead{$V\;$} & \colhead{$\sigma_V$} & \colhead{$B-V$} &
\colhead{$\sigma_{B-V}$} & \colhead{$U-B$} & \colhead{$\sigma_{U-B}$} &\colhead{
$N$} & \colhead{$N$} &
 \multicolumn{2}{c}{$E(B-V)$} \\
\colhead{} &
\multicolumn{2}{c}{1950} &
\multicolumn{2}{c}{2000} &
\colhead{} &
\colhead{} &
\colhead{} &
\colhead{} &
\colhead{} &
\colhead{} &
\colhead{} &
\colhead{} &
\colhead{$_{ni}$} &
\colhead{$_{ob}$} &
\colhead{S} &
\colhead{B} \\
\colhead{(1)} &
\colhead{(2)} &
\colhead{(3)} &
\colhead{(4)} &
\colhead{(5)} &
\colhead{(6)} &
\colhead{(7)} &
\colhead{(8)} &
\colhead{(9)} &
\colhead{(10)} &
\colhead{(11)} &
\colhead{(12)} &
\colhead{(13)} &
\colhead{(14)} &
\colhead{(15)} &
\colhead{(16)} &
\colhead{(17)}
}
\startdata
BS 16083$-$0172  & 14 48 59.5 & $+$48 57 33 &  14 50 41.6 & $+$48 45 14 &  84.2&  \phs 58.4 &  13.13 &   0.39 & \phs 0.61 &   0.11 &  \phs 0.09 & 0.01 &  2 &  4 & 0.03 &  0.01\\
BS 16542$-$0002  & 16 18 24.8 & $+$22 50 18 &  16 20 33.1 & $+$22 43 12 &  39.7&  \phs 42.7 &  12.62 &   0.08 & \phs 0.76 &   0.04 & $-$0.04    & 0.04 &  6 & 15 & 0.06 &  0.04\\
BS 16556$-$0081  & 16 42 37.8 & $+$31 32 02 &  16 44 32.8 & $+$31 26 34 &  52.6&  \phs 39.6 &  14.32 &   0.09 & \phs 0.48 &   0.08 & $-$0.25    & 0.02 &  3 &  6 & 0.03 &  0.02\\
BS 17569$-$0014  & 21 55 34.8 & $+$06 20 16 &  21 58 04.6 & $+$06 34 36 &  65.3& $-$36.2    &  12.99 &   0.11 & \phs 0.48 &   0.10 & $-$0.11    & 0.05 &  1 &  2 & 0.06 &  0.05\\
BS 17569$-$0023  & 21 57 07.8 & $+$05 54 18 &  21 59 37.9 & $+$06 08 41 &  65.2& $-$36.8    &  11.05 &   0.38 & \phs 1.44 &   0.03 &  \phs 1.39 & 0.04 &  2 &  4 & 0.05 &  0.04\\
BS 17569$-$0027  & 21 56 59.5 & $+$03 38 20 &  21 59 31.0 & $+$03 52 43 &  63.0& $-$38.2    &  13.98 &   0.13 & \phs 0.49 &   0.06 & $-$0.10    & 0.04 &  4 & 11 & 0.05 &  0.04\\
BS 17569$-$0028  & 21 58 10.1 & $+$03 26 17 &  22 00 41.7 & $+$03 40 43 &  63.0& $-$38.6    &  13.89 &   0.18 & \phs 0.56 &   0.06 & $-$0.17    & 0.03 &  4 &  8 & 0.05 &  0.03\\
BS 17569$-$0047  & 22 01 06.1 & $+$04 39 20 &  22 03 37.1 & $+$04 53 52 &  64.8& $-$38.4    &  14.37 &   0.10 & \phs 0.47 &   0.04 & $-$0.16    & 0.04 &  2 &  4 & 0.05 &  0.04\\
\enddata
\end{deluxetable}

\clearpage

\begin{deluxetable}{ccc}
\tabletypesize{\footnotesize}
\tablenum{4}
\tablecaption{Program Stars With Duplicate HK Survey Identifications \label{tbl$-$4}}
\tablewidth{0pt}
\tablehead{
\colhead{Cataloged Star Name} & \colhead{Alternate Name 1} & \colhead{Alternate Name 2}}
\startdata
BS 16034$-$0115 &  BS 16079$-$0069 & BS 16084$-$0001 \nl
BS 16076$-$0012 &  BS 16942$-$0037 & \nodata \nl                              
BS 16077$-$0077 &  BS 16081$-$0010 & \nodata \nl         
BS 16079$-$0133 &  BS 16084$-$0075 & \nodata \nl                                                            
BS 16082$-$0129 &  BS 16078$-$0085 & \nodata \nl                                                             
BS 16083$-$0043 &  BS 15620$-$0020 & \nodata \nl                                                             
CS 29502$-$0042 &  CS 29516$-$0041 & \nodata \nl                                                            
CS 31061$-$0032 &  CS 31063$-$0035 & \nodata \nl                                                            
CS 31061$-$0062 &  CS 31063$-$0069 & \nodata \nl                                                             
CS 31061$-$0072 &  CS 31063$-$0092 & \nodata \nl                                                             
CS 31079$-$0004 &  CS 31078$-$0002 & \nodata \nl                                                            
CS 31084$-$0022 &  CS 31083$-$0022 & \nodata \nl                                                           
\enddata
\end{deluxetable}%
   

\clearpage

\figcaption[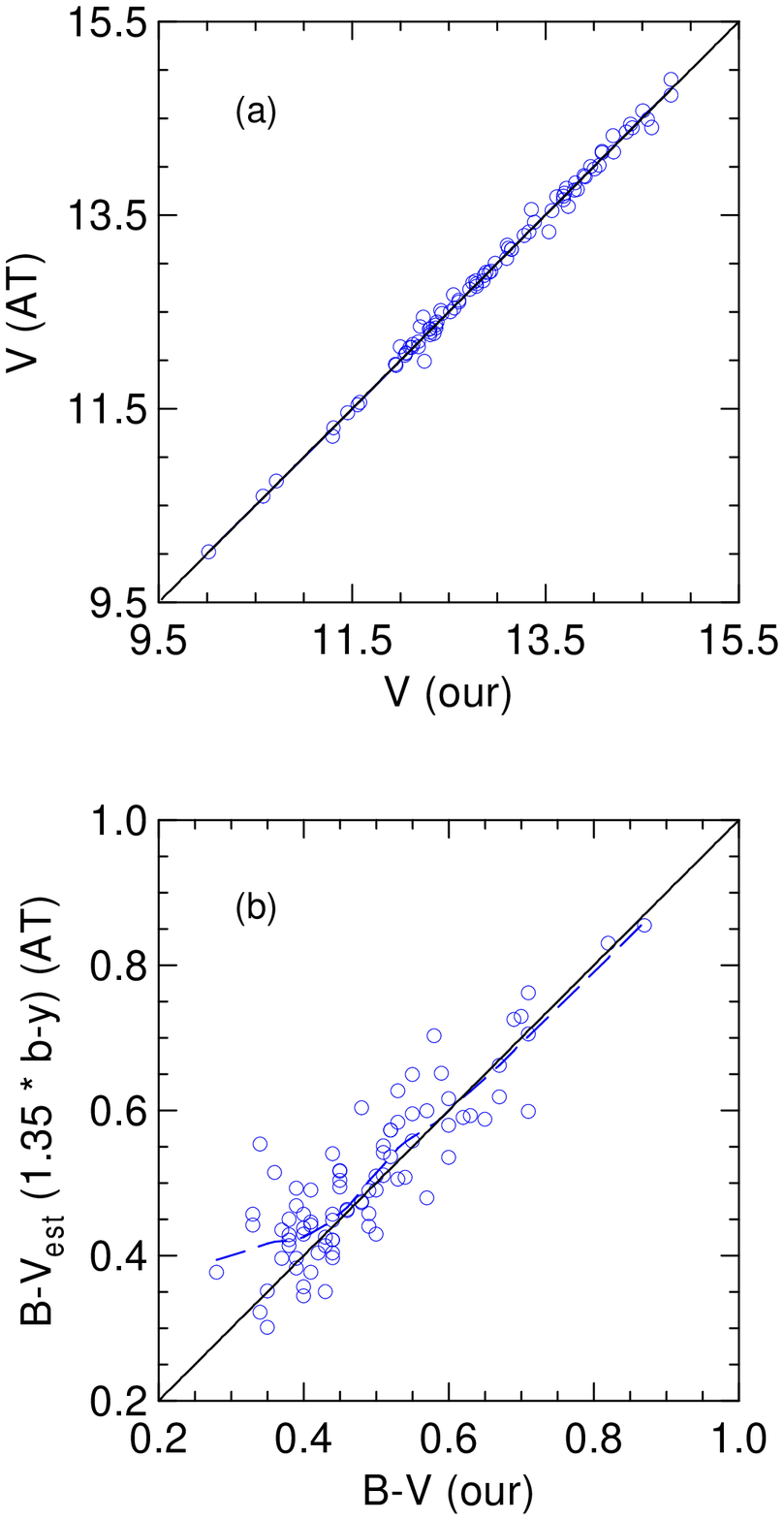]{(a) Comparison of the $V$ magnitudes for 85 stars in
common between the present sample, $V\; (our)$, and that of Anthony-Twarog et
al. (2000), $V\; (AT)$.  A one-to-one line is shown. (b) Comparison of the
measured, $(B-V) \; (our)$, and estimated, $(B-V)_{est} (AT)$, colors for the
same sample.  A one-to-one line is shown (solid line), along with a locally
weighted regression line (dashed line) that indicates a slight disagreement at
bluer colors. \label{fig1}}

\figcaption[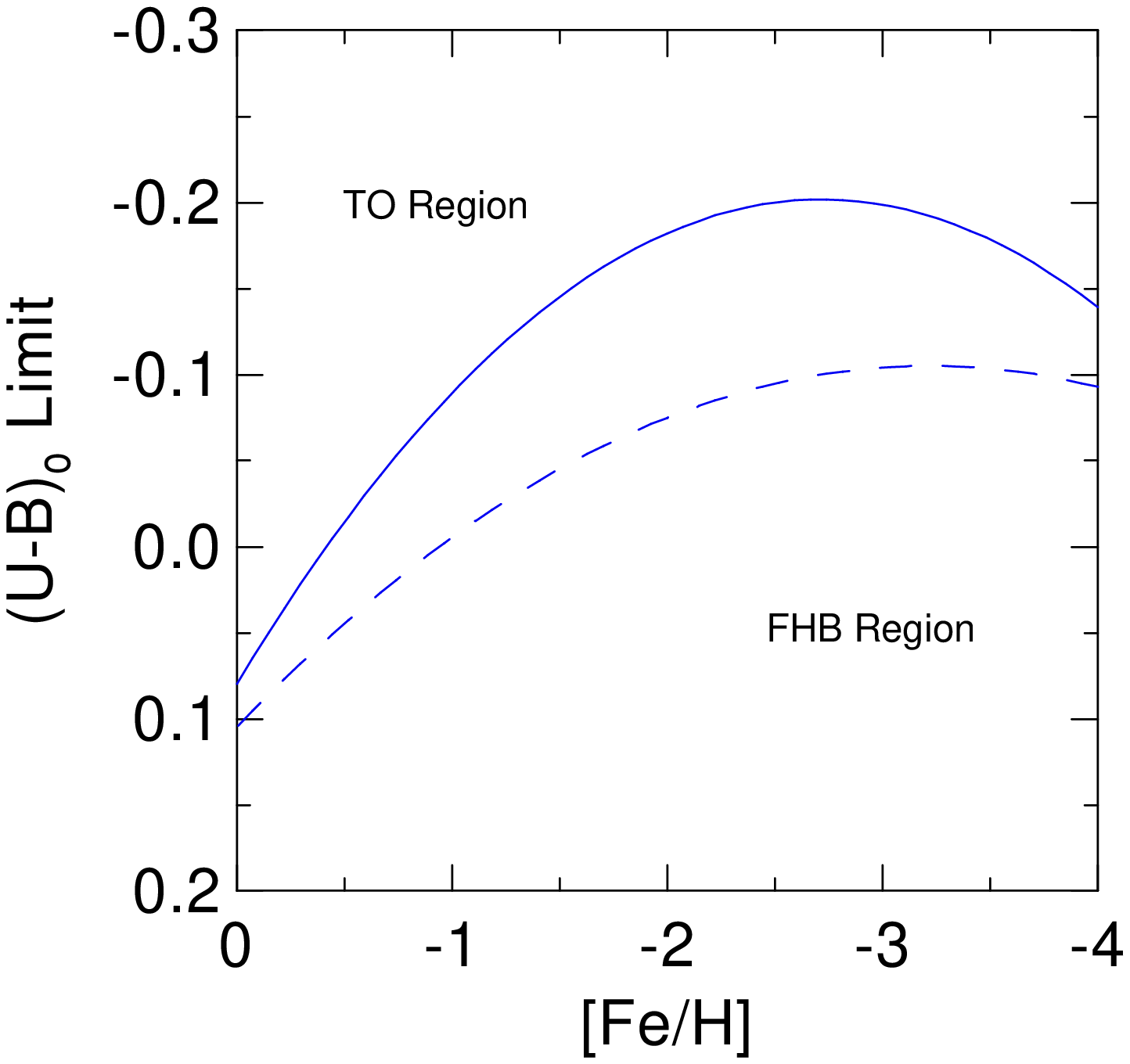]{Limiting $(U-B)_0$ colors of stars in the color range
$0.3 \le (B-V)_0 \le 0.5$, as a function of metallicity, obtained from the
Revised Yale Isochrones.  The region above the solid line corresponds to the
locations of stars with surface gravities of main-sequence TO stars, while the
region below the dashed line applies for stars with surface gravities of FHB
stars.  See text for additional explanation. \label{fig2}}

\figcaption[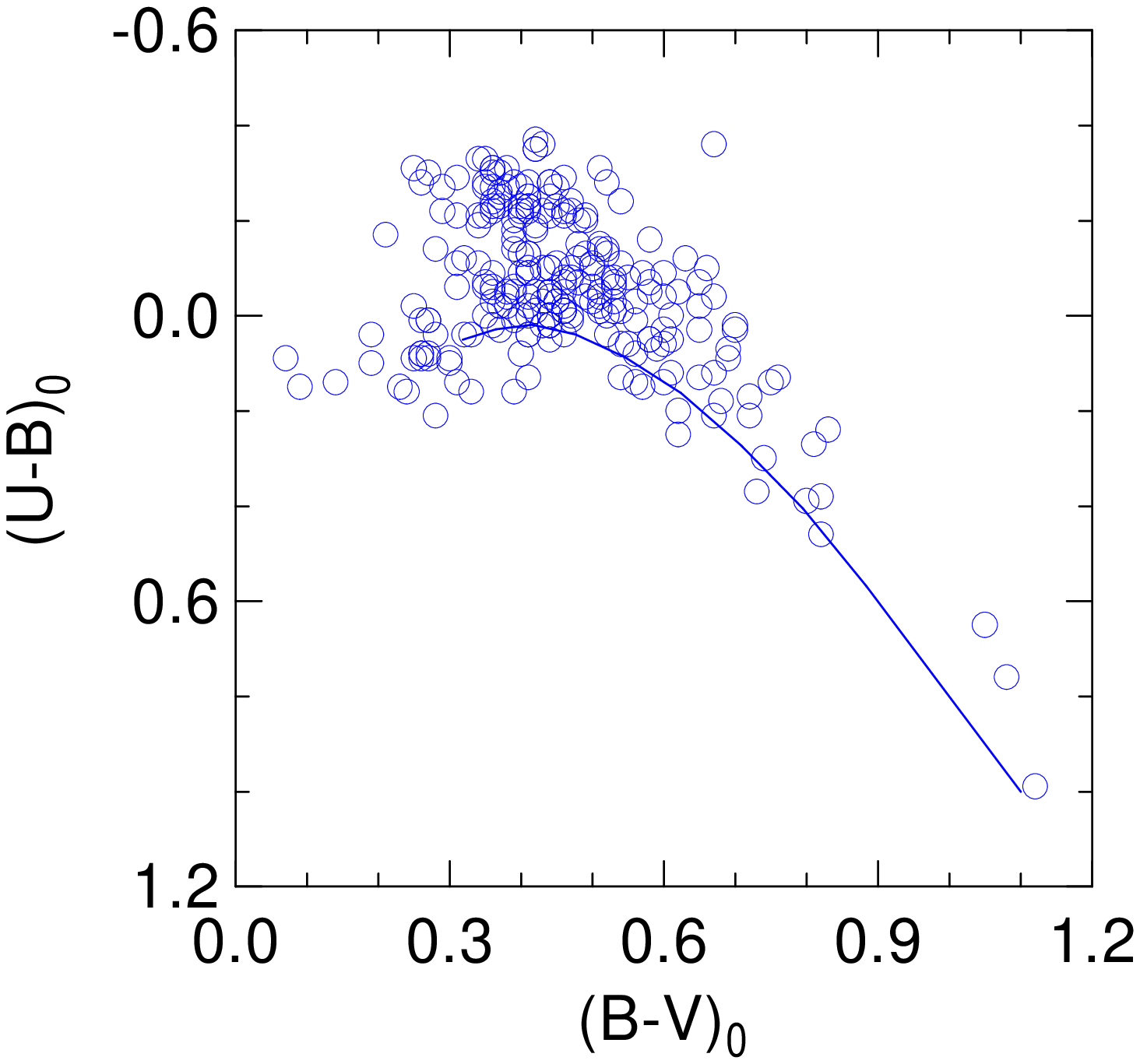]{De-reddened two-color diagram for the non-variable 
stars in our sample over the color range for which reliable metallicity
estimates can be obtained.  A solar-abundance main sequence line, obtained from
the Revised Yale Isochrones, is shown for comparison.
\label{fig3}}

\clearpage

\begin{figure}
\figurenum{1}
\plotone{fig1.eps}
\end{figure}

\clearpage

\begin{figure}
\figurenum{2}
\plotone{fig2.eps}
\end{figure}

\clearpage

\begin{figure}
\figurenum{3}
\plotone{fig3.eps}
\end{figure}

\end{document}